\documentclass[paper=a4,bibliography=totoc,numbers=noendperiod,parskip=full,headings=normal,draft=false]{scrartcl}

\usepackage[utf8]{inputenc}
\usepackage[T1]{fontenc}
\usepackage[ngerman,english]{babel}

\usepackage{amsmath}
\usepackage{amssymb,textcomp}
\usepackage{esint}
\usepackage{bbold}
\usepackage{graphicx}

\usepackage[affil-sl]{authblk}

\usepackage{lmodern}
\usepackage{microtype}

\DeclareMathOperator{\tr}{tr}
\DeclareMathOperator{\Sym}{\text{Sym}}

\newcommand{\skeina}{\parbox[m]{.7cm}{\includegraphics[width=.7cm]{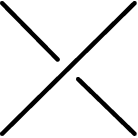}}} 
\newcommand{\skeinaa}{\parbox[m]{.7cm}{\includegraphics[width=.7cm, angle=90]{skein0}}}\newcommand{\skeinb}{\parbox[m]{.7cm}{\includegraphics[width=.7cm]{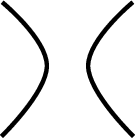}}} 
\newcommand{\skeinc}{\parbox[m]{.7cm}{\includegraphics[width=.7cm]{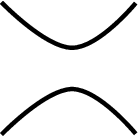}}} 
\newcommand{\skeinl}{\parbox[m]{1.1cm}{\includegraphics[width=1.1cm]{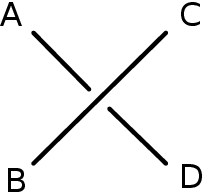}}} 

\newcommand{\skeinaarrows}{\parbox[m]{.7cm}{\includegraphics[width=.7cm]{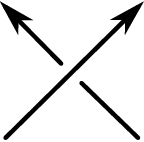}}} 
\newcommand{\skeinaarrowsmirror}{\parbox[m]{.7cm}{\includegraphics[width=.7cm]{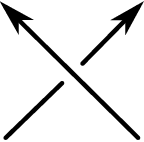}}}
\newcommand{\skeinbarrows}{\parbox[m]{.7cm}{\includegraphics[width=.7cm]{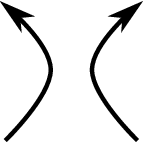}}}

\newcommand{\expec}[1]{\langle #1\rangle}

\usepackage{url}

\renewcommand{\S}[1]{\mathcal{S(}#1\mathcal{)}}
\newcommand{\U}[1]{\mathcal{U(}#1\mathcal{)}}

\begin{document}

\title{Extensions of the Duflo map and Chern-Simons expectation values}
\author[,a]{Hanno Sahlmann\thanks{email: hanno.sahlmann@gravity.fau.de}}
\author[,a]{Thomas Zilker\thanks{email: thomas.zilker@gravity.fau.de}}

\affil[a]{\normalsize Institute for Quantum Gravity, Friedrich-Alexander-Universität Erlangen-Nürnberg\\ Staudtstr. 7/B2, D-91058 Erlangen, Germany}
\date{\today}
\maketitle

\begin{abstract}
The Duflo map is a valuable tool for operator ordering in contexts in which the Kirillov-Kostant-Souriau bracket and its quantization plays a role. It has beautiful properties on the subspace of the symmetric algebra over a Lie algebra consisting of elements invariant under the adjoint action. In the present work, we focus on its action beyond this subspace: We calculate the image of the exponential map, to obtain a certain deformation of SU(2), and we discuss and compare modifications of its action on non-invariant elements. Also, an application to the calculation of Chern-Simons theory expectation values is discussed. 
\end{abstract}

%----------------------------------------
\section{Introduction}
%----------------------------------------
The Duflo-map  $Q_D$ \cite{Duflo:1977} 
\begin{equation}
Q_D: \S{\mathfrak{g}}\longrightarrow \U{\mathfrak{g}}
\end{equation}
is a map from the symmetric algebra $\S{\mathfrak{g}}$ over a Lie algebra $\mathfrak{g}$ into its universal enveloping algebra $\U{\mathfrak{g}}$. The symmetric algebra can be interpreted as consisting of polynomial functions on the dual $\mathfrak{g}^*$. One can think of 
$\mathfrak{g}^*$ as a phase space by equipping the symmetric algebra with a Poisson bracket, the  \emph{Kirillov-Kostant-Souriau bracket}. Then  $Q_D$ is a quantization map. Indeed, the Duflo map  is a special case of Kontsevich's quantization of Poisson manifolds \cite{Kontsevich:1997vb}. As such, it is a natural choice for a quantization map, whenever Lie algebras and those brackets play a role, and has consequently found its way into the physics literature. An elementary and intriguing example is the Duflo quantization of the hydrogen atom by Rosa and Vitale \cite{Rosa:2012pr} where they obtain the energy spectrum of the hydrogen atom (including the correct degeneracies). In that case, the use of the Duflo map makes the choice of an ordering in the Lenz-Runge vector obsolete. 

In loop quantum gravity, one of the variables is a (densitized) vector field $E$, taking values in su(2)*. The components of $E$ become non-commutative in the quantum theory. The non-commutativity is dictated by the structure constants of SU(2) and can be thought of as arising from a quantization of the Kirillov-Kostant-Souriau bracket. This makes the Duflo map interesting in the context of loop quantum gravity. Indeed, it has found several applications in this field. The first was a proposal for the quantization of the area operator \cite{Alekseev:2000hf}. In this case the use of the Duflo map leads to a different area spectrum. Further applications include a quantization of 2+1 dimensional gravity with LQG methods \cite{Noui:2011im,Pranzetti:2014xva,Cianfrani:2016ogm}\footnote{It appears that in this application, a different, though closely related, quantization map is used -- see the discussion in section \ref{sec:NouiComparison}.} and a momentum representation for LQG \cite{Guedes:2013vi}.

More recently, the Duflo map has also been used to quantize a version (using the traces of surface holonomies) of the boundary condition satisfied by spherically symmetric isolated horizons in the SU(2) formalism. This approach lead to the occurrence of Chern-Simons theory expectation values \cite{Sahlmann:2011rv,Sahlmann:2011uh} without bringing in Chern-Simons theory by hand as is done in the standard treatments of black holes in LQG \cite{Smolin1995,Ashtekar2000,Domagala2004,Kaul1998,Kaul2000,Engle2009,Engle2010,Beetle:2010rd}.

From a mathematical point of view, the property that gives the Duflo map an edge over symmetric quantization is that it is an isomorphism of algebras when restricted to a certain subalgebra, namely the subalgebra $\S{\mathfrak{g}}^{\mathfrak{g}}$ of elements invariant under the adjoint action. With the no table exceptions of \cite{Noui:2011im,Guedes:2013vi}, all the applications of the Duflo map to physics mentioned above only made use of the Duflo map on this subalgebra. The main goal of the present work is to study the Duflo map on the larger space $\S{\mathfrak{g}}$, for the case $G$ = SU(2).

Since the action of $Q_D$ away from $\S{\mathfrak{g}}^{\mathfrak{g}}$ is arguably less fixed\footnote{It seems that it can be fixed by requiring functoriality, but we are not aware of a proof of that statement. For a given $G$, the action of $Q_D$ away from $\S{\mathfrak{g}}^{\mathfrak{g}}$ can be changed.}, we will also consider different extensions $Q'_D$. 

Explicit results for the action of $Q_D$ (and some modified versions) on the whole space  $\S{\mathfrak{g}}$  can be used to generalize \cite{Sahlmann:2011uh,Sahlmann:2011rv} to expectation values of (untraced) holonomies. While this is interesting in its own right, it is also a step towards finding a preferred ordering for a quantization of (an exponentiated version of) the isolated horizon boundary condition in LQG. Such a generalization involves in particular finding an extension of the Duflo map to terms which are not gauge invariant, since the results of \cite{Sahlmann:2011uh} provide evidence that the Duflo map yields the correct ordering for the products of flux operators occurring in the series expansion of the surface ordered exponential.\footnote{It turns out that the boundary conditions imply that the canonical variables used in the description form a 2-connection in the language of higher gauge theory \cite{Zilker:2017aey}. Consequently, the Duflo map is effectively used to quantize a surface holonomy in higher gauge theory for a particular choice of 2-group.} In a first attempt we will apply the explicit formula for the Duflo map used in \cite{Sahlmann:2011uh} also to terms that are not gauge invariant. 
In particular, we will extend \cite{Sahlmann:2011rv,Sahlmann:2011uh} to the full (untraced) SU(2) isolated horizon boundary condition.

As we will also be computing what one could call the \emph{quantization of the exponential map}, the result will be an operator valued matrix that can be understood as a quantum deformation of SU(2). A comparison to other quantum deformations, such as SU$_q$(2) will appear elsewhere \cite{h+t-inprep}. 

The paper is organized as follows: We will start with a short introduction to the Duflo map and its applications in LQG, in particular the one in \cite{Sahlmann:2011uh}. Next, we will evaluate the Duflo map on the particular type of non-gauge-invariant terms relevant for the current considerations. This section will also contain a precise definition of the Duflo map. 
In section \ref{sec:NouiComparison}, we will compare the result of our calculations to results obtained using a different continuation (to non-gauge-invariant terms) of the Duflo map \cite{Noui:2011aa}. Finally, in section \ref{sec:CSExpValues} we will use our results to quantize a specific function of flux operators (essentially their exponential) using the Duflo map and elaborate on the possibility to derive skein relations from this. A section devoted to the discussion of our findings as well as a brief outlook will conclude this paper.

%-----------------------------------------------------
\section{An introduction to the Duflo map}
%-----------------------------------------------------
The Duflo-map \cite{Duflo:1977}, a generalization of the Harish-Chandra isomorphism \cite{Harishchandra:1951}, is a map from the symmetric algebra $\S{\mathfrak{g}}$ over a Lie algebra $\mathfrak{g}$ to its universal enveloping algebra $\U{\mathfrak{g}}$, with marvelous properties. More precisely, it is an algebra isomorphism 
\begin{equation}
\label{eq:sub}
Q_D: \S{\mathfrak{g}}^{\mathfrak{g}}\longrightarrow Z(\U{\mathfrak{g}})
\end{equation}
between the subalgebra of invariant elements of $\S{\mathfrak{g}}$ and the center of $\U{\mathfrak{g}}$. Sometimes, the name Duflo map is reserved to the algebra isomorphism \eqref{eq:sub} above, but we will use it for the action on all of $\S{\mathfrak{g}}$, as it was originally defined by Duflo. 

In \cite{Alekseev:2000hf}, it was observed that $Q_D$ can be used as an ordering prescription for invariant functions on a Lie group that preserves all the classical relations. The Poisson structure that is quantized is given by the Kirillov-Kostant-Souriau (KKS) bracket which defines a Poisson structure on $\S{\mathfrak{g}}$. To this end we regard $\S{\mathfrak{g}}$ as polynomial functions over $\mathfrak{g}^*$. Given $a\in\mathfrak{g}$, the corresponding function is $E_a(z)= -i z(a)$, and the bracket reads  

\begin{equation}
\label{eq:KKS}
\{E_a(z),E_b(z)\}:=E_{[a,b]}(z).
\end{equation}
We note that this is not the only possible choice for a Poisson structure on $\S{\mathfrak{g}}$. More discussion on this topic will appear elsewhere \cite{h+t-inprep}. 

The explicit formula for the Duflo map $Q_{D}$ on $\S{\mathfrak{g}}^{\mathfrak{g}}$ as given in \cite{Duflo:1977} is
\begin{equation}
Q_{D} = Q_{S} \circ j^{\frac{1}{2}} \left( \partial \right),
\label{eqn:DufloMap}
\end{equation}
where $Q_{S}$ denotes symmetric quantization (i.e. the Poincaré-Birkhoff-Witt isomorphism) and $j^{\frac{1}{2}} \left( \partial \right)$ is an infinite order differential operator obtained from inserting the natural derivative $\partial^i$ on $\Sym(\mathfrak{g})$ into the function 
\begin{equation}
j^{\frac{1}{2}} (x) = \sqrt{\operatorname{det} \left( \frac{\operatorname{sinh} \frac{\operatorname{ad}_{x}}{2}}{\frac{\operatorname{ad}_{x}}{2}} \right)} 
\end{equation}
on $\mathfrak{g}$, with $\operatorname{ad}_{x}$ denoting the adjoint action of $x$. 

Both, the KKS bracket \eqref{eq:KKS} and the Duflo map \eqref{eqn:DufloMap} are defined on all of $\S{\mathfrak{g}}$. The distinguishing feature of $Q_D$ (the algebra isomorphism property), however, only holds on the subalgebra $\S{\mathfrak{g}}^{\mathfrak{g}}$, whence other extensions  of this map to the full symmetric algebra are conceivable. For instance, one might consider $\S{\mathfrak{g}}$ as an $\S{\mathfrak{g}}^{\mathfrak{g}}$-module and continue $Q_D$ as a morphism of $\S{\mathfrak{g}}^{\mathfrak{g}}$-modules. This raises the question whether there is an extension to all of $\S{\mathfrak{g}}$ which is in some sense more natural than any other. One aim of this work is to consider the merits of several such extensions.

Let us make these things more explicit and also introduce some notation. An element $x$ of $\mathfrak{g}$ can be written as 
\begin{equation}
x=x^iT_i
\end{equation}
with $T_i$ a basis of $\mathfrak{g}$. Then we can introduce a basis $X^i$ of $\mathfrak{g}^*$ with $X^i(x)=x^i$ and a basis $F_i$ of $\mathfrak{g}^{**}$ by $F_i(X^j)=\delta_i^j$. 

Defining functions $E_j$ on $\mathfrak{g}^*$ via 
\begin{equation}
E_j=-iF_j, 
\end{equation}
we can identify polynomials in the $E_j$ with elements of the symmetric algebra $\S{\mathfrak{g}}$. With respect to this identification, $E_i$ corresponds to the generator $T_i$ of $\mathfrak{g}$. 
The derivative $\partial^i$ is given by 
\begin{equation}
\partial^iE_j=\delta^i_j. 
\end{equation} 
and extended to polynomials by linearity and Leibniz rule. 

A typical element of the universal enveloping algebra $U(\mathfrak{g})$ is given by linear combinations of monomials
\begin{equation}
\widehat{E}_{i_1}\widehat{E}_{i_2} \ldots \widehat{E}_{i_n}
\end{equation}
where $[\widehat{E}_{i},\widehat{E}_{j}]=f_{ij}^{~\,k}\widehat{E}_k$ and the $f_{ij}^{~\,k}$ are the structure constants of $\mathfrak{g}$. 

Finally, we can also write down the ingredients of $Q_D$: The function $j^{\frac{1}{2}} \left( \partial \right)$ is obtained by the replacement $x\rightarrow T_i\partial^i$. 
Symmetric quantization is given by 
\begin{equation}
Q_S(E_{i_1}E_{i_2} \ldots E_{i_n})=\widehat{E}_{(i_1}\widehat{E}_{i_2} \ldots \widehat{E}_{i_n)}. 
\end{equation}

\section{Evaluating the Duflo map for SU(2)}
\label{sec:DufloCalc}
In this section we will evaluate the Duflo map on the particular type of non-gauge-invariant terms relevant for the application to horizons in LQG.
For a complementary approach to the calculation of the image of non-gauge-invariant terms under the Duflo map see \cite{Guedes:2013vi}. 

For the case of SU(2) the function $j^{\frac{1}{2}} (x)$ can be evaluated explicitly as follows:
\begin{equation}
\begin{split}
j^{\frac{1}{2}} (x) &= \sqrt{\operatorname{det} \left( \sum_{N=0}^{\infty} \frac{1}{(2N+1)!} \left[ \frac{\operatorname{ad}_{x}}{2} \right]^{2N} \right)} \\
&= \sqrt{\left( \sum_{N=0}^{\infty} \frac{1}{(2N+1)!} \left[ \frac{-|x|^{2}}{4} \right]^{N} \right)^{2}} \\ &= \left| \sum_{N=0}^{\infty} \frac{1}{(2N+1)!} \left[ \frac{||x||^{2}}{8} \right]^{N} \right| \\
&= \sum_{N=0}^{\infty} \frac{1}{(2N+1)!} \frac{1}{8^{N}} \kappa_{m_1 n_1} \dots \kappa_{m_N n_N} x^{m_1} x^{n_1} \dots x^{m_N} x^{n_N}
\end{split}
\end{equation}
with $\kappa_{mn} = -2\,\delta_{mn}$ denoting the components of the Killing metric on $\mathfrak{su}(2)$.  Here, we used the series expansion of $\operatorname{sinh}$ in the first line and the fact that the determinant is given by the product of the eigenvalues in the second line. The eigenvalues of $\operatorname{ad}_{x}^{2}$ are given by $0$ and $-|x|^{2}$, where the latter occurs with a multiplicity of two. We then made use of the relationship $|x|^{2} := \delta_{mn} x^{m} x^{n} = - \frac{1}{2} \kappa_{mn} x^{m} x^{n} =: -\frac{1}{2} ||x||^{2}$ and, in the last line, we also used that the series between the absolute value signs corresponds to the function $\frac{\operatorname{sinh}(y)}{y}$, which is positive everywhere and thus allows us to drop the absolute value. \\
Now we want to extend this map from terms of the form $||E||^{2n}$ to terms of the form $||E||^{2n} E_{i}$. The latter are not gauge invariant and hence the Duflo map does not act as an algebra isomorphism on them. We will thus have to evaluate the Duflo map -- as given by equation (\ref{eqn:DufloMap}) -- explicitly for such terms. We start by computing the action of the differential operator $j^{\frac{1}{2}} (\partial)$ on this type of terms. Since $||\partial||^{2n} = \left[ || \partial||^{2} \right]^{n}$, we first calculate (for $k \geq 1$)

\begin{equation}
\begin{split}
|| \partial||^{2} \left[ ||E||^{2k} E_{i} \right] &= \kappa_{mn} \kappa^{i_{1}i_{2}} \dots \kappa^{i_{2k-1}i_{2k}} \partial^{m} \partial^{n} E_{i_{1}} \dots E_{i_{2k}} E_{i} \\ &= \kappa_{mn} \kappa^{i_{1}i_{2}} \dots \kappa^{i_{2k-1}i_{2k}} \partial^{m} \left[ \left( 2k \right) \delta^{n}_{i_{1}} E_{i_{2}} \dots E_{i_{2k}} E_{i} + \delta^{n}_{i} E_{i_{1}} \dots E_{i_{2k}} \right] \\ &= \kappa_{mn} \kappa^{i_{1}i_{2}} \dots \kappa^{i_{2k-1}i_{2k}} \left( 2k \right) \delta^{n}_{i_{1}} \delta^{m}_{i_{2}} E_{i_{3}} \dots E_{i_{2k}} E_{i} \\ &+ \kappa_{mn} \kappa^{i_{1}i_{2}} \dots \kappa^{i_{2k-1}i_{2k}} \left( 2k \right) \delta^{n}_{i_{1}} \left( 2k-2 \right) \delta^{m}_{i_{3}} E_{i_{2}} E_{i_{4}} \dots E_{i_{2k}} E_{i} \\ &+ \kappa_{mn} \kappa^{i_{1}i_{2}} \dots \kappa^{i_{2k-1}i_{2k}} \left( 2k \right) \delta^{n}_{i_{1}} \delta^{m}_{i} E_{i_{2}} \dots E_{i_{2k}} \\ &+ \kappa_{mn} \kappa^{i_{1}i_{2}} \dots \kappa^{i_{2k-1}i_{2k}} \delta^{n}_{i} \left( 2k \right) \delta^{m}_{i_{1}} E_{i_{2}} \dots E_{i_{2k}} \\ &= \left( 2k+3 \right) \left( 2k \right) ||E||^{2(k-1)} E_{i}
\end{split}
\end{equation}

whence we obtain for $n \leq k$

\begin{equation}
\begin{split}
|| \partial||^{2n} \left[ ||E||^{2k} E_{i} \right] &= \prod_{m=k-n+1}^{k} \left( 2m+3 \right) \left( 2m \right) ||E||^{2(k-n)} E_{i} \\ &= \frac{(2k+1)!}{(2k-2n+1)!} \frac{2k+3}{2k-2n+3} ||E||^{2(k-n)} E_{i}
\end{split}
\end{equation}

and $|| \partial||^{2n} \left[ ||E||^{2k} E_{i} \right] = 0$ for $n > k$. The result of the action of the infinite order differential operator $j^{\frac{1}{2}} (\partial)$ on this particular type of terms is thus given by

\begin{equation}
j^{\frac{1}{2}} (\partial) \left[ ||E||^{2k} E_{i} \right] = \sum_{N=0}^{k} \frac{1}{(2N+1)!} \frac{1}{8^{N}} \frac{(2k+1)!}{(2k-2N+1)!} \frac{2k+3}{2k-2N+3} ||E||^{2(k-N)} E_{i}.
\label{eqn:DiffOp}
\end{equation}

Since $Q_{S}$ is a linear map, in order to compute $Q_{D}$ we need the action of $Q_{S}$ again on terms of the form $||E||^{2n} E_{i}$. However, we can reduce this to the problem of calculating $Q_{S}(||E||^{2(n+1)})$ via

\begin{equation}
\begin{split}
Q_{S} (||E||^{2(k+1)}) &= \kappa^{i_{1} i_{2}} \dots \kappa^{i_{2k+1} i_{2k+2}} \hat{E}_{(i_{1}} \dots \hat{E}_{i_{2k+2})}\\
&= \kappa^{i_{1} i_{2}} \dots \kappa^{i_{2k+1} i_{2k+2}} \frac{1}{2k+2} \sum_{l=1}^{2k+2} \hat{E}_{(i_{1}} \dots \hat{E}_{i_{l-1}} \hat{E}_{i_{l+1}} \dots \hat{E}_{i_{2k+2})} \hat{E}_{i_{l}}\\
&= \kappa^{i_{1} i_{2}} \dots \kappa^{i_{2k+1} i_{2k+2}} \frac{1}{2k+2} \sum_{l=1}^{2k+2} \hat{E}_{(i_{1}} \dots \hat{E}_{i_{2k+1})} \hat{E}_{i_{2k+2}}\\
&= \kappa^{i_{1} i_{2}} \dots \kappa^{i_{2k+1} i_{2k+2}} \hat{E}_{(i_{1}} \dots \hat{E}_{i_{2k+1})} \hat{E}_{i_{2k+2}}\\ &= Q_{S} (||E||^{2k} E_{i_{2k+1}}) \kappa^{i_{2k+1} i_{2k+2}} \hat{E}_{i_{2k+2}},
\end{split}
\end{equation}

where $\hat{E_{i}} := Q_{S}(E_{i})$ and we used the definition of $Q_{S}$ in the first line and the definition of symmetrization in the second line. In the third line, we then relabelled the dummy indices $i_{l}$ and $i_{2k+2}$ in each term of the sum and used the total symmetry of the first $2k+1$ indices to restore the original kappas in front of the sum. Thus, in the fourth line, all terms in the sum are the same and we can express the result in terms of $Q_{S} (||E||^{2n} E_{i_{2k+1}})$. Since $Q_{S} (||E||^{2n} E_{i_{2k+1}})$ has to be proportional to $\hat{E}_{i_{2k+1}}$, we can thus write $Q_{S} (||E||^{2n} E_{i_{2k+1}})$ in terms of $Q_{S} (||E||^{2(k+1)})$ as

\begin{equation}
Q_{S} (||E||^{2n} E_{i_{2k+1}}) = \frac{Q_{S} (||E||^{2(n+1)})}{\Delta_{\mathfrak{su}(2)}} \hat{E}_{i_{2k+1}},
\label{eqn:NGIviaGI}
\end{equation}

where $\Delta_{\mathfrak{su}(2)} := \kappa^{ij} \hat{E}_{i} \hat{E}_{j}$ denotes the generator of the center of $\U{\mathfrak{su}(2)}$. Now, we only need to evaluate $Q_{S}$ on terms of the form $||E||^{2n}$, which is given in \cite{Kirillov:1999} as

\begin{equation}
Q_{S}(r^{2n}) = \frac{(-1)^{n-1}}{4^{n}} \sum_{k=0}^{n} \binom{2n+1}{2k} B_{2k} \left( 4^{k} - 2 \right) \left( 1 - 4C \right)^{n-k}
\end{equation}

with $r=|E|$, $C=Q_{S}(r^{2})$ and $B_{2k}$ denoting the $2k$-th Bernoulli number. In our notation, this formula thus reads

\begin{equation}
Q_{S}(||E||^{2k}) = - \frac{1}{8^{k}} \sum_{m=0}^{k} \binom{2k+1}{2m} B_{2m} \left( 2^{2m} - 2 \right) \left( 1 + 8 \Delta_{\mathfrak{su}(2)} \right)^{k-m}.
\label{eqn:SYM}
\end{equation}

The combination of equations (\ref{eqn:DiffOp}), (\ref{eqn:NGIviaGI}) and (\ref{eqn:SYM}) leads to a rather lengthy expression. In the spin-\textonehalf-representation, however, things simplify drastically. More precisely, we have $\Pi^{(1/2)}(\hat{E}_{i}) = \tau_{i} = - \frac{i}{2} \sigma_{i}$, with $\sigma_{i}$ denoting the Pauli matrices, and therefore we obtain $\Pi^{(1/2)} (\Delta_{\mathfrak{su}(2)}) = \kappa^{ij} \tau_{i} \tau_{j} = \frac{3}{8}$, which leads to 

\begin{equation}
\begin{split}
\Pi^{(1/2)}(Q_{S}(||E||^{2k})) &= -\frac{1}{8^{k}} \sum_{m=0}^{k} \binom{2k+1}{2m} B_{2m} \left( 2^{2k} - 2^{2k-2m+1} \right)\\ &= -\frac{1}{2^{k}} \sum_{m=0}^{2k} \binom{2k+1}{m} B_{m} + \frac{1}{8^{k}} \sum_{m=0}^{2k} \binom{2k+1}{m} B_{m} 2^{2k-m+1}\\ &= -\frac{1}{2^{k}} \delta_{k,0} + \frac{1}{8^{k}} \left( 2k+1 \right) \sum_{m=0}^{2k} \binom{2k}{m} B_{m} \frac{2^{2k-m+1}}{2k-m+1}\\ &= - \delta_{k,0} + \frac{1}{8^{k}} \left( 2k+1 \right) \left( 1 + \delta_{k,0} \right) = \frac{1}{8^{k}} \left( 2k+1 \right),
\end{split}
\end{equation}

where we used a basic property of Bernoulli numbers to get the third line and a specific version of Faulhaber's formula in the first equality of the last line. Note also that the terms with $m=1$ from the two sums in the second line cancel each other. Hence, it does not matter whether we use Bernoulli numbers of first or second kind. For all other odd $m$, the corresponding terms in the sums vanish, since in this case $B_{m} = 0$. Now, inserting this result, together with equations (\ref{eqn:DiffOp}) and (\ref{eqn:NGIviaGI}), into equation (\ref{eqn:DufloMap}) we finally get

\begin{equation}
\begin{split}
\Pi^{(1/2)}&(Q_{D}(||E||^{2k} E_{i})) =\\ &= \sum_{N=0}^{k} \frac{1}{(2N+1)!} \frac{1}{8^{N}} \frac{2k+3}{2k-2N+3} \frac{(2k+1)!}{(2k-2N+1)!} \Pi^{(1/2)}(Q_{S}(||E||^{2(k-N)} E_{i}))\\ &= \sum_{N=0}^{k} \frac{1}{(2N+1)!} \frac{1}{8^{N}} \frac{2k+3}{2k-2N+3} \frac{(2k+1)!}{(2k-2N+1)!} \frac{\frac{1}{8^{k-N+1}} \left( 2(k-N+1) + 1 \right)}{\frac{3}{8}} \tau_{i}\\ &= \frac{1}{3\cdot 8^{k}} \sum_{N=0}^{k} \left( 2k+3 \right) \frac{(2k+1)!}{(2N+1)!(2k-2N+1)!} \tau_{i}\\ &= \frac{\frac{2}{3}k + 1}{8^{k}} \sum_{N=0}^{k} \frac{1}{2k+2} \binom{2k+2}{2N+1} \tau_{i}\\ &= \frac{1}{2^{k}} \frac{\frac{2}{3}k+1}{k+1} \tau_{i}.
\end{split}
\label{eqn:Duflo_oddPower}
\end{equation}

For the sake of completeness we also provide the image of terms of the form $||E||^{2k}$ under the Duflo map, which is given by

\begin{equation}
Q_{D}(||E||^{2k}) = \left[ Q_{D}(||E||^{2}) \right]^{k} = \left[ \Delta_{\mathfrak{su}(2)} + \frac{1}{8}  \mathbb{1} \right]^{k}
\end{equation}

and in the spin-\textonehalf-representation simplifies to

\begin{equation}
\Pi^{(1/2)}(Q_{D}(||E||^{2k})) = \frac{1}{2^{k}}.
\label{eqn:Duflo_evenPower}
\end{equation}

\section{Comparison of different extensions}
\label{sec:NouiComparison}

In this section we want to compare the Duflo map to a different (but related) choice of quantization map used by the authors of \cite{Noui:2011im}. They claim to use the Duflo map themselves, but they state a slightly different formula for it (seemingly copied from the arXiv version of \cite{Alekseev:2000hf}). Their formula has the advantage that in their context (non-commutative holonomies in (2+1)-gravity) it produced a very appealing result, namely a relation to knot theory via Kauffman's bracket. Since the difference in the formulas for the Duflo map crucially influences the result, in the following we will apply our formula -- based on the one given in \cite{Sahlmann:2011uh, Duflo:1977} -- to their calculation and compare the results afterwards. The complete expression Noui et al. consider in \cite{Noui:2011im} is rather complicated and fortunately not needed here. The relevant part consists of terms of the form

\begin{equation}
\frac{z^{p}}{p!} \tau^{i_{1}} \dots \tau^{i_{p}} \otimes Q(E_{i_{1}} \dots E_{i_{p}}),
\end{equation}

where $z$ is some (purely imaginary) constant, $\tau^{i} = -\frac{i}{2} \sigma^{i}$ denote the generators of $\mathfrak{su}(2)$ in the spin-\textonehalf-representation and $Q: \S{\mathfrak{g}} \longrightarrow \U{\mathfrak{g}}$ denotes some quantization map, i.e. in our case either $Q_{S}$ or $Q_{D}$.
Since the domain of the map $Q$ is $\S{\mathfrak{g}}$, we know that $Q(E_{i_{1}} \dots E_{i_{p}})$ has to be symmetric in all indices and hence we can equivalently write

\begin{equation}
\begin{split}
&\frac{z^{p}}{p!} \tau^{(i_{1}} \dots \tau^{i_{p})} \otimes Q(E_{i_{1}} \dots E_{i_{p}}) \\
= &\frac{z^{p}}{p!} \begin{cases} \frac{1}{2^{k}} \{ \tau^{i_{1}}, \tau^{i_{2}} \} \dots \{ \tau^{i_{2k-1}}, \tau^{i_{2k}} \} \otimes Q(E_{i_{1}} \dots E_{i_{2k}}) & \text{if } p = 2k\\ \frac{1}{2^{k}} \{ \tau^{i_{1}}, \tau^{i_{2}} \} \dots \{ \tau^{i_{2k-1}}, \tau^{i_{2k}} \} \tau^{i_{2k+1}} \otimes Q(E_{i_{1}} \dots E_{i_{2k+1}}) & \text{if } p = 2k+1 \end{cases}\\
= &\frac{z^{p}}{p!} \begin{cases} \frac{1}{2^{k}} \kappa^{i_{1}i_{2}} \dots \kappa^{i_{2k-1}i_{2k}} \otimes Q(E_{i_{1}} \dots E_{i_{2k}}) & \text{if } p = 2k\\ \frac{1}{2^{k}} \kappa^{i_{1}i_{2}} \dots \kappa^{i_{2k-1}i_{2k}} \tau^{i_{2k+1}} \otimes Q(E_{i_{1}} \dots E_{i_{2k+1}}) & \text{if } p = 2k+1 \end{cases}\\
= &\frac{z^{p}}{p!} \begin{cases} \frac{1}{2^{k}} \mathbb{1} \otimes Q(||E||^{2k}) & \text{if } p = 2k\\ \frac{1}{2^{k}} \tau^{i_{2k+1}} \otimes Q(||E||^{2k} E_{i_{2k+1}})& \text{if } p = 2k+1 \end{cases}
\end{split}
\end{equation}

Hence, we are down to evaluating $Q$ on the type of terms we already considered in the previous section. Since we are working in the spin-\textonehalf-representation here, we can use equations (\ref{eqn:Duflo_oddPower}) and (\ref{eqn:Duflo_evenPower}) from the previous section to obtain

\begin{equation}
\begin{split}
\sum_{p=0}^{\infty} &\frac{z^{p}}{p!} \tau^{i_{1}} \dots \tau^{i_{p}} \otimes Q_{D} (E_{i_{1}} \dots E_{i_{p}}) = \\ &= \sum_{k=0}^{\infty} \left[ \frac{z^{2k}}{(2k)!} \frac{1}{2^{2k}} \mathbb{1} \otimes \mathbb{1} + \frac{z^{2k+1}}{(2k+1)!} \frac{1}{2^{2k}} \frac{2}{3} \frac{2k+3}{2k+2} \tau^{i} \otimes \tau_{i} \right] \\ &= \operatorname{cosh} \left( \frac{z}{2} \right) \mathbb{1} \otimes \mathbb{1} + \frac{4}{3} \left[ \operatorname{sinh} \left( \frac{z}{2} \right) + \frac{\operatorname{cosh} \left( \frac{z}{2} \right) - 1}{\frac{z}{2}} \right] \tau^{i} \otimes \tau_{i} \\ &= \operatorname{cos} \left( \frac{o \hbar \lambda}{2} \right) \mathbb{1} \otimes \mathbb{1} - \frac{4i}{3} \left[ \operatorname{sin} \left( \frac{o \hbar \lambda}{2} \right) + \frac{\operatorname{cos} \left( \frac{o \hbar \lambda}{2} \right) - 1}{\frac{o \hbar \lambda}{2}} \right] \tau^{i} \otimes \tau_{i} \, ,
\end{split}
\end{equation}

where we substituted $z$ with its explicit form in the last line.
This expression is more complicated than the one obtained by the authors of \cite{Noui:2011im} and doesn't lead to their appealing result. Since both our and their version of the Duflo map appear in the literature, it may be illuminating to further investigate the two formulas. The difference seems to have its foundation in different formulas for $j^{\frac{1}{2}} \left( x \right)$, which is given by

\begin{equation}
j^{\frac{1}{2}} \left( x \right) = \sqrt{\operatorname{det} \left( \frac{\operatorname{sinh} \frac{\operatorname{ad}_{x}}{2}}{\frac{\operatorname{ad}_{x}}{2}} \right)} 
\end{equation}

in \cite{Duflo:1977}, whereas the authors of \cite{Alekseev:2000hf} seem to use

\begin{equation}
\tilde{j}^{\frac{1}{2}} \left( x \right) = \operatorname{det} \left( \frac{\operatorname{sin} \operatorname{ad}_{x}}{\operatorname{ad}_{x}} \right) 
\end{equation}

instead (at least this formula is stated in the arXiv version of their paper). However, at this stage, the precise nature of the difference between $Q_{D}$ and $\tilde{Q}_{D}^{NPP}$ when applied to non-gauge-invariant terms remains unclear.

\section{Quantized exponential map}
\label{sec:CSExpValues}

As we will explain in section \ref{se_cs}, for the applications we have in mind, the element  
\begin{equation}
\label{eq_qexp}
Q \left[ \operatorname{exp} \left( - \frac{8\pi i}{k} \kappa^{ij} E_{i} T_{j} \right) \right]
\end{equation}
in $\U{\mathfrak{su}(2)}$ is of particular importance. Note that this can be understood as a ``quantization of the exponential map'', so it might also be interesting from a purely mathematical point of view. 

We will calculate \eqref{eq_qexp} using the various extensions proposed in the previous text. For comparison we will give the results for $Q=Q_{D}$, $Q=\tilde{Q}_{D}$, $Q=\tilde{Q}^{NPP}_{D}$ and $Q=Q_{S}$, where $Q_{D}$ denotes the Duflo map as defined in eqn. (\ref{eqn:DufloMap}), $Q_{S}$ denotes symmetric quantization as above, $Q=\tilde{Q}_{D}$ coincides with $Q_{D}$ on terms of the form $||E||^{2n}$ but is continued via 
\begin{equation}
\tilde{Q}_{D} (||E||^{2n} E_{i}) = Q_{D} (||E||^{2n}) Q_{D} (E_{i}),
\end{equation}
and $\tilde{Q}^{NPP}_{D}$ is the Duflo map used by the authors of \cite{Noui:2011im}, i.e. $\tilde{Q}^{NPP}_{D} (||E||^{2n}) = \frac{1}{8^{n}}$ and is continued in the same way as $\tilde{Q}_{D}$.

As a first step, we expand the exponential as a series yielding

\begin{equation}
\begin{split}
&\operatorname{exp} \left( - \frac{8\pi i}{k} \kappa^{ij} E_{i} T_{j} \right) = \operatorname{cos} \left( \frac{4\pi}{\sqrt{2}k} ||E|| \right) \mathbb{1}_{2} - \frac{8\pi i}{k} \frac{\operatorname{sin} \left( \frac{4\pi}{\sqrt{2}k} ||E|| \right)}{\frac{4\pi}{\sqrt{2}k} ||E||} \kappa^{ij} E_{i} T_{j} \\ &= \sum_{m=0}^{\infty} \frac{(-1)^{m} \left( \frac{4\pi}{\sqrt{2}k} \right)^{2m}}{(2m)!} ||E||^{2m} \mathbb{1}_{2} - \frac{8\pi i}{k} \sum_{m=0}^{\infty} \frac{(-1)^{m} \left( \frac{4\pi}{\sqrt{2}k} \right)^{2m}}{(2m+1)!} ||E||^{2m} \kappa^{ij} E_{i} T_{j}.
\end{split}
\label{eq:expE}
\end{equation}

We will now consider the application of the different quantization maps to \eqref{eq:expE}. In order to allow for an easier comparison, we will also express our results in the basis used in \cite{Noui:2011im}.

For $Q = Q_{D}$ we obtain

\begin{equation}
\begin{split}
Q_{D} &\left[ \operatorname{exp} \left( - \frac{8\pi i}{k} \kappa^{ij} E_{i} \left( T_{j} \right)^{A}_{~D} \right) \right]^{C}_{~B} \\ &= \operatorname{cos} \left( \frac{2\pi}{k} \right) \delta^{A}_{~D} \delta^{C}_{~B} + \frac{4i}{3} \left[ \operatorname{sin} \left( \frac{2\pi}{k} \right) + \frac{1 - \operatorname{cos} \left( \frac{2\pi}{k} \right)}{\frac{2\pi}{k}} \right] \sum_{i} \left( T_{i} \right)^{A}_{~D} \left( T_{i} \right)^{C}_{~B} \\ &= \left[ \operatorname{cos} \left( \frac{2\pi}{k} \right) - \frac{i}{3} \operatorname{sin} \left( \frac{2\pi}{k} \right) - \frac{i}{3} \frac{1 - \operatorname{cos} \left( \frac{2\pi}{k} \right)}{\frac{2\pi}{k}} \right] \delta^{A}_{~B} \delta^{C}_{~D} \\ &- \left[ \operatorname{cos} \left( \frac{2\pi}{k} \right) + \frac{i}{3} \operatorname{sin} \left( \frac{2\pi}{k} \right) + \frac{i}{3} \frac{1 - \operatorname{cos} \left( \frac{2\pi}{k} \right)}{\frac{2\pi}{k}} \right] \epsilon^{AC} \epsilon_{BD} \, .
\end{split}
\end{equation}

\newpage
In the case $Q = \tilde{Q}_{D}$ we have

\begin{equation}
\begin{split}
\tilde{Q}_{D} &\left[ \operatorname{exp} \left( - \frac{8\pi i}{k} \kappa^{ij} E_{i} \left( T_{j} \right)^{A}_{~D} \right) \right]^{C}_{~B} \\ &= \operatorname{cos} \left( \frac{2\pi}{k} \right) \delta^{A}_{~D} \delta^{C}_{~B} + 2i \operatorname{sin} \left( \frac{2\pi}{k} \right) \sum_{i} \left( T_{i} \right)^{A}_{~D} \left( T_{i} \right)^{C}_{~B} \\ &= \left[ \operatorname{cos} \left( \frac{2\pi}{k} \right) - \frac{i}{2} \operatorname{sin} \left( \frac{2\pi}{k} \right) \right] \delta^{A}_{~B} \delta^{C}_{~D} - \left[ \operatorname{cos} \left( \frac{2\pi}{k} \right) + \frac{i}{2} \operatorname{sin} \left( \frac{2\pi}{k} \right) \right] \epsilon^{AC} \epsilon_{BD} \, .
\end{split}
\end{equation}

Using $Q = \tilde{Q}^{NPP}_{D}$ we are left with

\begin{equation}
\begin{split}
\tilde{Q}^{NPP}_{D} &\left[ \operatorname{exp} \left( - \frac{8\pi i}{k} \kappa^{ij} E_{i} \left( T_{j} \right)^{A}_{~D} \right) \right]^{C}_{~B} \\ &= \operatorname{cos} \left( \frac{\pi}{k} \right) \delta^{A}_{~D} \delta^{C}_{~B} + 4i \operatorname{sin} \left( \frac{\pi}{k} \right) \sum_{i} \left( T_{i} \right)^{A}_{~D} \left( T_{i} \right)^{C}_{~B} \\ &= \left[ \operatorname{cos} \left( \frac{\pi}{k} \right) - i \operatorname{sin} \left( \frac{\pi}{k} \right) \right] \delta^{A}_{~B} \delta^{C}_{~D} - \left[ \operatorname{cos} \left( \frac{\pi}{k} \right) + i \operatorname{sin} \left( \frac{\pi}{k} \right) \right] \epsilon^{AC} \epsilon_{BD} \\ &= \operatorname{e}^{-\frac{i\pi}{k}} \delta^{A}_{~B} \delta^{C}_{~D} - \operatorname{e}^{\frac{i\pi}{k}} \epsilon^{AC} \epsilon_{BD} \, .
\end{split}
\end{equation}

The choice $Q = Q_{S}$ results in

\begin{equation}
\begin{split}
Q_{S} &\left[ \operatorname{exp} \left( - \frac{8\pi i}{k} \kappa^{ij} E_{i} \left( T_{j} \right)^{A}_{~D} \right) \right]^{C}_{~B} \\ &= \left[ \operatorname{cos} \left( \frac{\pi}{k} \right) - \frac{\pi}{k} \operatorname{sin} \left( \frac{\pi}{k} \right) \right] \delta^{A}_{~D} \delta^{C}_{~B} + \frac{4 i}{3} \left[ 2 \operatorname{sin} \left( \frac{\pi}{k} \right) + \frac{\pi}{k} \operatorname{cos} \left( \frac{\pi}{k} \right) \right] \sum_{i} \left( T_{i} \right)^{A}_{~D} \left( T_{i} \right)^{C}_{~B} \\ &= \left[ \operatorname{cos} \left( \frac{\pi}{k} \right) - \frac{2i}{3} \operatorname{sin} \left( \frac{\pi}{k} \right) \right] \delta^{A}_{~B} \delta^{C}_{~D} - \left[ \operatorname{cos} \left( \frac{\pi}{k} \right) + \frac{2i}{3} \operatorname{sin} \left( \frac{\pi}{k} \right) \right] \epsilon^{AC} \epsilon_{BD} \\ &- \frac{i\pi}{3k} \left[ \operatorname{cos} \left( \frac{\pi}{k} \right) - 3i \operatorname{sin} \left( \frac{\pi}{k} \right) \right] \delta^{A}_{~B} \delta^{C}_{~D} + \frac{i\pi}{3k} \left[ \operatorname{cos} \left( \frac{\pi}{k} \right) + 3i \operatorname{sin} \left( \frac{\pi}{k} \right) \right] \epsilon^{AC} \epsilon_{BD} \, .
\end{split}
\end{equation}

Lastly, let us also consider $Q = \tilde{Q}_{S}$, which denotes the continuation of $Q_{S}$ analogous to $\tilde{Q}_{D}$. The expression then reads

\begin{equation}
\begin{split}
\tilde{Q}_{S} &\left[ \operatorname{exp} \left( - \frac{8\pi i}{k} \kappa^{ij} E_{i} \left( T_{j} \right)^{A}_{~D} \right) \right]^{C}_{~B} \\ &= \left[ \operatorname{cos} \left( \frac{\pi}{k} \right) - \frac{\pi}{k} \operatorname{sin} \left( \frac{\pi}{k} \right) \right] \delta^{A}_{~D} \delta^{C}_{~B} + \frac{4\pi i}{k} \operatorname{cos} \left( \frac{\pi}{k} \right) \sum_{i} \left( T_{i} \right)^{A}_{~D} \left( T_{i} \right)^{C}_{~B} \\ &= \left[ \operatorname{cos} \left( \frac{\pi}{k} \right) - \frac{i\pi}{k} \operatorname{e}^{-\frac{i\pi}{k}} \right] \delta^{A}_{~B} \delta^{C}_{~D} - \left[ \operatorname{cos} \left( \frac{\pi}{k} \right) + \frac{i\pi}{k} \operatorname{e}^{\frac{i\pi}{k}} \right] \epsilon^{AC} \epsilon_{BD} \, .
\end{split}
\end{equation}

Since the Duflo map is supposed to be a deformed version of symmetric quantization, it is interesting to note that, while $\tilde{Q}^{NPP}_{D}$ and $Q_{S}$ both produce $\frac{\pi}{k}$ as argument of the occuring $\operatorname{sin}$ and $\operatorname{cos}$ functions, our version yields $\frac{2\pi}{k}$ instead. Additionally, the fact that $\tilde{Q}^{NPP}_{D}$ leads to a similarly simple result as in \cite{Noui:2011im} indicates that $\tilde{Q}^{NPP}_{D}$ might be the best choice to define an ordering for the quantization of products of flux operators.

%--------------------------------------------------------------------------------------------
\section{Application to quantum Chern Simons theory and black holes}
\label{se_cs}
%-------------------------------------------------------------------------------------------
In loop quantum gravity, one of the variables is a (densitized) vector field $E$, taking values in su(2)*, 
\begin{equation}
E(x)=E_i^a(x)\;\partial_a\otimes t^i.
\end{equation}
Upon quantization, due to the use of Lie algebra-valued variables as configuration variables, the momenta do not commute any more. Choosing a certain regularization\footnote{Details will appear elsewhere \cite{h+t-inprep}.} one can decompose 
\begin{equation}
\widehat{E}_i^a(x)=\widehat{E}^a(x)\widehat{E}_i(x)
\end{equation}
where the $\widehat{E}^a(x)$ denote certain operator valued distributions that commute among each other, and 
\begin{equation}
[\widehat{E}_i(x), \widehat{E}_j(y)]=\delta_{x,y}\epsilon_{ijk}\widehat{E}_k(x)
\end{equation}
The non-commutativity is dictated by the structure constants of SU(2) and can be thought of as arising from a quantization of the KKS bracket \eqref{eq:KKS}. 

There are two direct applications of the results of the calculation of the quantized exponential map in the context of LQG. Both center around the operators 
\begin{equation}
\mathbf{W_{S}} := \left[ \mathcal{P} \operatorname{exp} \varoiint_{S} \left( \frac{2\pi}{ick} 
h^{-1}T_ih\; \kappa^{ij} \epsilon_{abc}\;\widehat{E}_j^a \;  \text{d}x^b\text{d}x^c\right) \right].
\end{equation}
The integral is a surface ordered exponential integral (for details see for example \cite{Arefeva:1980,Sahlmann:2011uh}).  The holonomy $h$ connects the points of $S$ with a base point on $\partial S$, $(T_i)$ is a basis of the Lie algebra su(2), $\kappa$ denotes its Cartan-Killing metric and $c$ is a constant that depends on the application. 

Since the $\widehat{E}_i^a(x)$ are non-commutative as explained above, ${W_{S}}$ is not well defined as it stands. $Q_D$ can be used as a quantization map, and makes it a well defined operator on holonomy functionals 
\begin{equation}
h_e[A]=\mathcal{P} \operatorname{exp} \oint_{e} A_a\;  \text{d}x^a. 
\end{equation}
In the case where $\mathbf{W_S}$ acts on a holonomy $h_e$, with $e$ and $S$ having only a single transversal intersection, the action of $\mathbf{W_S}$ is given by the quantization of the exponential map. 

One application of the operators $\mathbf{W_S}$ is in the context of the treatment of black holes in LQG, \cite{Smolin:1995vq,Ashtekar:1997yu,Engle:2009vc}. In \cite{Sahlmann:2011xu,Sahlmann:2011rv} it was sketched how the $\mathbf{W_S}$ could be used to determine the structure of the surface Hilbert space for a black hole horizon in LQG. In that application, all surfaces $S$ are lying within the horizon. This application will be discussed in greater detail elsewhere \cite{h+t-inprep}.

In the following we will discuss the application \cite{Sahlmann:2011uh,Sahlmann:2011rv} of $\mathbf{W_S}$ to the calculation of Chern-Simons (CS) expectation values of Wilson loops, using structures in the kinematical quantization of LQG.  The approach consists of two crucial steps: First one uses the fact that under the CS path integral the curvature of the connection can be replaced by a functional derivative with respect to the connection. Secondly, one applies a non-Abelian version of Stokes' theorem \cite{Arefeva:1980} to identify holonomies of the connection with surface ordered exponentials of the curvature. Combining these two steps one can thus calculate CS expectation values of traces of holonomies via
\begin{equation}
\langle \tr h_{\partial S}[A] \rangle = \int_{\overline{\mathcal{A}}} \left( \operatorname{tr} \mathbf{W_{S}} \right) \, [\operatorname{exp}(iS_{CS}[A])] \mu_{AL}[A],
\end{equation}
for a suitable constant $c$ in the definition of $\mathbf{W_S}$.  
With this method it was shown how to recover the known CS expectation values for the unknot and the Hopf link and for gauge groups SU(2) and SU(3). 

However, the expectation values were calculated in a piecemeal fashion, turning one loop $\partial S$ into an operator $\mathbf{W_S}$ at a time, and calculating its action under the path integral. Now that we have extensions of the Duflo map at our disposal, we can aim for skein relations among the expectation values. The argument goes as follows.  

We consider the path integral expectation value of the traces of holonomies along the components of a link $L$:

\begin{equation}
\langle F_L \rangle_\text{CS} = \int_{\overline{\mathcal{A}}}\operatorname{exp}(iS_{CS}[A])\;F_L[A] \;\text{d}\mu_{AL}[A].
\end{equation}

\begin{figure}[h]
\centerline{\includegraphics[width=.7\textwidth]{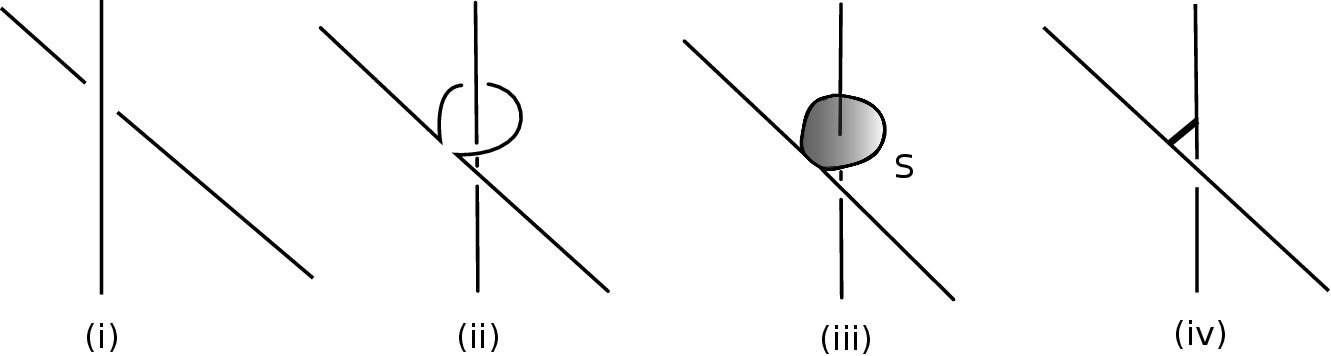}}
\caption{\label{fi_crossing}  Manipulation of a crossing of two holonomy strands, using the operators $\mathbf{W_S}$}
\end{figure}

Consider two holonomy strands passing each other as in figure \ref{fi_crossing}(i). As the expectation value does not depend on smooth deformations of $L$, we can deform the one strand in the manner shown in fig.\ \ref{fi_crossing}(ii).\footnote{Strictly speaking, the deformation depicted in (ii) is smooth only so long as the circle around the other holonomy strand does not get closed completely. If it is not completely closed, however, the replacement in step (iii) (see below) is only an approximation. This approximation can be made arbitrarily good, classically, and we will assume in the following that this is also true in the quantum theory.} By applying the non-abelian Stokes theorem (for details see \cite{Sahlmann:2011uh}), we can replace the curved section of the deformed strand by a certain ordered exponential integral $\mathcal{I}_S$ of the curvature of $A$ over a surface $S$ bounded by the curved section, 

\begin{equation}
\langle F_L \rangle_\text{CS} = \int_{\overline{\mathcal{A}}}\operatorname{exp}(iS_{CS}[A])\;(\widetilde{F}_L)^I{}_J[A] \;(\mathcal{I}_S)^J{}_I[A]\;
\text{d}\mu_{AL}[A],
\end{equation}

see also (iii).  $(\widetilde{F}_L)^I{}_J$ is obtained from the original functional by removing the holonomy along $\partial S$. In the next step,  $\mathcal{I}_S$ can be replaced by a functional differential operator acting on the action term. For the action 

\begin{equation}
S_\text{CS}=\frac{k}{4\pi}\int_M\tr(A\wedge dA+\frac{2}{3}A\wedge A\wedge A))
\end{equation}

it holds that 

\begin{equation}
\label{eq_fder}
\frac{\delta}{\delta A^{j}(x)} e^{iS_{\text{CS}}}[A]= \frac{ick}{2\pi}\;\kappa_{jl}\; F^{l}(x)e^{iS_{\text{CS}}}[A]. 
\end{equation}

$c$ is a Lie algebra dependent constant ($c=1/4$ for $A$ an SU(2) connection). Thus 

\begin{equation}
\langle F_L \rangle_\text{CS} = \int_{\overline{\mathcal{A}}}  (\mathbf{W_S})^J{}_I\left[\operatorname{exp}(iS_{CS}[A])\right]\;(\widetilde{F}_L)^I{}_J[A] \;
\text{d}\mu_{AL}[A].
\end{equation}

$\mathbf{W_S}$ is an operator obtained from $\mathcal{I}_S$ by substituting \eqref{eq_fder}. These functional derivatives can be rigorously defined and they do not commute with each other. Hence an ordering is needed and is provided by (an extension of) the Duflo map. 

In the next step, partial functional integration gives

\begin{equation}
\langle F_L \rangle_\text{CS} = \int_{\overline{\mathcal{A}}}  \left[\operatorname{exp}(iS_{CS}[A])\right]\;(\mathbf{W_S}^\dagger)^J{}_I\left[(\widetilde{F}_L)^I{}_J[A]\right] \;
\text{d}\mu_{AL}[A].
\end{equation}

It turns out that $\mathbf{W_S}^\dagger$ acts only at intersection points of $S$ with holonomy loops
\cite{Sahlmann:2011uh}. In the situation at hand, there is only one intersection. In that case, the action is given by inserting the ``quantized exponential map'' into the remaining holonomy strand, 

\begin{equation}
(\mathbf{W_S}^\dagger)^A{}_D  \left[(\widetilde{F}_L)^D{}_A[A]\right]  =
Q \left[ \operatorname{exp} \left( - \frac{8\pi i}{k} \kappa^{ij} E_i T_j {}^A{}_D  \right) \right]^C_{~B}
(\widetilde{F}_L)^D{}_A{}^B{}_C[A].
\end{equation}

The added pair of indices on $\widetilde{F}_L$ is due to the fact that a strand was cut at the intersection point with the surface $S$. This leads to a coupling between the two strands by an intertwiner, as in (iv) of fig.\ \ref{fi_crossing}. In graphical notation, we can write 

\begin{equation}
Q \left[ \operatorname{exp} \left( - \frac{8\pi i}{k} \kappa^{ij} E_{i} \left( T_{j} \right)^{A}_{~D} \right) \right]^{C}_{~B}\widehat{=}\; \skeinl .
\end{equation}

Then, expanding the resulting intertwiner in a suitable basis, we obtain an expression that can be compared to the skein relation of knot invariants. 

In section \ref{sec:CSExpValues}, we had calculated the quantized exponential map in the spin-\textonehalf-re\-pre\-sen\-ta\-tion. The space of intertwiners in this case is 2 dimensional. There are two relevant bases, 

\begin{equation}\label{eq_basis1}
\skeinb \;\widehat{=}\; \delta^{A}_{~B} \delta^{C}_{~D}, \qquad \skeinc\; \widehat{=} -\;\epsilon^{AC} \epsilon_{BD}
\end{equation}

and 

\begin{equation}\label{eq_basis2}
\skeinb \;\widehat{=}\; \delta^{A}_{~B} \delta^{C}_{~D}, \qquad \skeinaa\; \widehat{=} \; \delta^{A}_{~D} \delta^{C}_{~B}. 
\end{equation}

They are adapted for comparison to the skein relations for the Kauffman bracket

\begin{equation}
\skeina = A \;\skeinb + A^{-1}\;\skeinc
\label{eqn:Kauffman}
\end{equation}

and the Jones polynomial

\begin{equation}
-t^{-1}\;\skeinaarrows\;+(t^{\frac{1}{2}}-t^{-\frac{1}{2}})\;\skeinbarrows\; +t\;\skeinaarrowsmirror\;=0.
\end{equation}

It is well known since \cite{Witten:1988hf} that the CS expectation values are closely related to both invariants. While the expectation values are framing dependent, the invariants are not. The Jones polynomial is obtained from the expectation values in standard framing\footnote{Standard framing is the framing obtained from the consideration of a Seifert surface for the link.}, while the bracket contains an additional factor with the writhe as an exponent, making it a regular isotopy invariant. Let us introduce the shortcut

\begin{equation}
\exp E := \operatorname{exp} \left( - \frac{8\pi i}{k} \kappa^{ij} E_{i} T_{j} \right).
\end{equation}

When $\widetilde{Q}^{NPP}_D$ is expanded in basis \eqref{eq_basis1}, we obtain

\begin{equation}
\widetilde{Q}^{NPP}_D \left[ \exp E \right] =\operatorname{e}^{-\frac{i\pi}{k}} \;\skeinb\;+ \operatorname{e}^{\frac{i\pi}{k}}\;\skeinc\end{equation}

whereas  $Q_D$ gives

\begin{equation}
\begin{split}
Q_D[\exp E]&= \left[ \operatorname{cos} \left( \frac{2\pi}{k} \right) - \frac{i}{3} \operatorname{sin} \left( \frac{2\pi}{k} \right) - \frac{i}{3} \frac{1 - \operatorname{cos} \left( \frac{2\pi}{k} \right)}{\frac{2\pi}{k}} \right] \skeinb \\ &\qquad\qquad+ \left[ \operatorname{cos} \left( \frac{2\pi}{k} \right) + \frac{i}{3} \operatorname{sin} \left( \frac{2\pi}{k} \right) + \frac{i}{3} \frac{1 - \operatorname{cos} \left( \frac{2\pi}{k} \right)}{\frac{2\pi}{k}} \right] \skeinc.
\end{split}
\end{equation}

Using

\begin{equation}\label{eq_recoupling}
\skeinc\; = \;\skeinaa\; -\;\skeinb
\end{equation}

we can also expand these expressions in basis \eqref{eq_basis2}:

\begin{equation}
\widetilde{Q}^{NPP}_D \left[ \exp(E) \right] =\left(\operatorname{e}^{-\frac{i\pi}{k}}-\operatorname{e}^{\frac{i\pi}{k}}\right)
 \;\skeinb\;+\operatorname{e}^{\frac{i\pi}{k}}\;\skeinaa
\end{equation}

and 

\begin{equation}
\label{eq_qd}
\begin{split}
Q_D[\exp E]&= -\frac{2i}{3}\left[   \operatorname{sin} \left( \frac{2\pi}{k} \right) + \frac{1 - \operatorname{cos} \left( \frac{2\pi}{k} \right)}{\frac{2\pi}{k}} \right] \skeinb \\ &\qquad\qquad+\left[ \operatorname{cos} \left( \frac{2\pi}{k} \right) + \frac{i}{3} \operatorname{sin} \left( \frac{2\pi}{k} \right) + \frac{i}{3} \frac{1 - \operatorname{cos} \left( \frac{2\pi}{k} \right)}{\frac{2\pi}{k}} \right] \skeinaa.
\end{split}
\end{equation}
It is clear and remarkable that $\widetilde{Q}^{NPP}_D$ reproduces the skein relation of the Kauffman bracket with 

\begin{equation}
A=\operatorname{e}^{\frac{i\pi}{k}}.
\end{equation}
 
It is equally clear that ${Q}_D$ gives no direct relationship to either the Jones polynomial or the Kauffman bracket. 

\begin{figure}[h]
\centerline{\includegraphics[width=.35\textwidth]{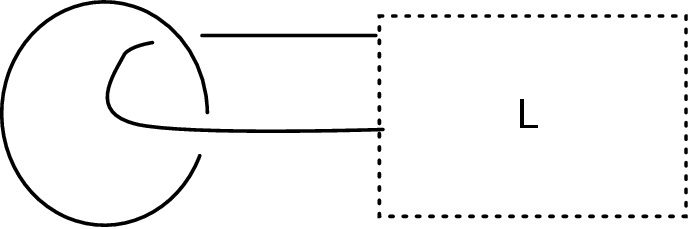}}
\caption{\label{fi_link} Linking an unknot with a link $L$}
\end{figure}

On the other hand, it was shown in \cite{Sahlmann:2011uh,Sahlmann:2011rv} that ${Q}_D$ successfully reproduces the relation (see figure \ref{fi_link})
\begin{equation}
\expec{\bigcirc \cup L}= q^{-\frac{3}{2}}(q+q^{-1})\expec{L}
\end{equation}
for the Jones polynomial, if one uses the relation 
\begin{equation}
\bigcirc= \tr \left({Q}_D[\exp E]\right)
\end{equation}
directly, without recourse to skein relations, whereas one can see from the results presented above, that this is not the case for $\widetilde{Q}^{NPP}_D$. 

Since the CS expectation values are known to be framing dependent, one might wonder how this can affect the results for the skein relations above. Indeed, the introduction of the surface $S$ in the calculation above endows part of the link with a specific framing. Taking into account this framing and the difference to standard framing, one would have to multiply the terms on the right hand side of \eqref{eq_qd} with $\Delta$ and $\Delta^2$, respectively, where \cite{Witten:1988hf}

\begin{equation}
\Delta:= q^{-\frac{3}{4}}. 
\end{equation}

However, this does not solve the problem of interpreting \eqref{eq_qd} in terms of a standard skein relation. 

\section{Conclusion \& Outlook}
\label{sec:Conclusion}
In the present work, we have considered different extensions of the Duflo map to $\S{\mathfrak{g}}$, as well as some variants of the Duflo map. Explicit calculations have been given for $\mathfrak{g}=\mathfrak{su}(2)$, in particular the image of the element 
\begin{equation}
 \operatorname{exp} \left( - \frac{8\pi i}{k} \kappa^{ij} E_{i} T_{j} \right)
\end{equation}
in the spin-\textonehalf-representation. 

Interpreting the Duflo map as a quantization map, we have applied it and its variants to the calculation of CS expectation values according to a prescription detailed in \cite{Sahlmann:2011uh,Sahlmann:2011rv}. The results are very interesting, but not straightforward to interpret:
Using the variant $\widetilde{Q}^{NPP}_D$ of \cite{Noui:2011aa} one can reproduce the skein relation of the Kauffman bracket. This was already observed in \cite{Noui:2011aa}. The calculation we have presented here is in a substantially different setting though, and thus serves to emphasize the importance and versatility of $\widetilde{Q}^{NPP}_D$ as a quantization map. 

Somewhat surprisingly, $Q_D$ does not seem to be able to reproduce the skein relation with the path integral method presented here. This is in contrast to the results \cite{Sahlmann:2011uh,Sahlmann:2011rv} that show that certain relations among CS expectation values can be reproduced correctly by $Q_D$. Those same calculations seem to fail, however, for $\widetilde{Q}^{NPP}_D$.

We can only speculate about the reasons for these incongruent results. One reason might be that we are missing something in the translation between the mathematical results of the Duflo map (section \ref{sec:NouiComparison}) and the CS expectation values. Another potential source of problems is the fact that we are using the classical recoupling theory in equations like \eqref{eq_recoupling}, where the recoupling theory of $U_q(\mathfrak{su}(2))$ might be expected. It would also be desirable to understand better what distinguishes  $\widetilde{Q}^{NPP}_D$ mathematically, and how it is related to $Q_D$. 

Indeed there is another way to approach a very similar problem. In \cite{Pranzetti:2014xva}, the author considers the commutator of the constraints of 2+1 gravity in a kinematical quantization based on \cite{Noui:2011aa}, and in particular on the use of $\widetilde{Q}^{NPP}_D$. It turns out that requiring anomaly freeness of these commutators fixes the expectation value of the spin-\textonehalf~unknot to the corresponding quantum group dimension. As a consequence, the recoupling theory of the quantum group applies, and more complicated expectation values take values consistent with other approaches. As these constraints could be viewed as playing a role analogous to the quantum IH boundary conditions in our work, an analysis of their commutation relations could shed further light on the situation considered in the present work. This is currently under investigation \cite{h+t-inprep}. 

It is intriguing that the structure  relevant for the application to CS theory takes the form $Q[\exp(E)]$, i.e., a quantization of the exponential map. We would like to further analyze what kind of deformation of SU(2) this object might represent. These questions, as well as an application of $Q_D$ to the quantum theory of spherically symmetric isolated horizons will also be considered elsewhere \cite{h+t-inprep}.

As a final remark, let us note that the IH boundary conditions imply that the canonical variables used in the description form a 2-connection in the sense of higher gauge theory \cite{Zilker:2017aey}. Consequently, we have used the Duflo map to quantize a surface holonomy in higher gauge theory for a particular choice of 2-group. Thus the present work might be a useful starting point when quantizing higher gauge theory. 

\section*{Acknowledgments}
We thank the referee for helpful comments on the manuscript. 
This work was supported by the Emerging Fields Project \emph{Quantum Geometry} of the Friedrich-Alexander-Universit\"at Erlangen-N\"urnberg (FAU). TZ was supported by the Elite Network of the State of Bavaria.

\bibliographystyle{utphys}
\bibliography{bibliography_final}
\end{document}